\documentclass[a4]{JHEP3}
\usepackage{graphicx}
\def\micro{{\tt micrOMEGAs1.2}~}
\def\isajet{{\tt Isajet7.69}~}
\def\suspect{{\tt Suspect2.2}~}
\def\softsusy{{\tt SOFTSUSY1.8.3}~}
\def\spheno{{\tt SPHENO2.20}~}
\def\mneuto{m_{\tilde{\chi}_1^0}}
\def\stauo{\tilde{\tau}_1}

\def\mstauo{m_{\stauo}}

\def\m0{M_0}
\def\mhf{M_{1/2}}
\def\omegah{$\Omega h^2$}

\preprint{LAPTH-1026/04 \\ hep-ph/0402161 \\}
\title{Uncertainties in Relic Density Calculations in mSUGRA\label{dm}}
\author{B.C. Allanach$^1$, G.~B\'elanger$^1$, F.~Boudjema$^1$,  A. Pukhov$^2$,
W. Porod$^3$ \\
1. LAPTH, 9 Chemin de Bellevue, F-74941 Annecy-le-Vieux, France\\
2. Skobeltsyn Institute of Nuclear Physics, Moscow State University, Moscow, Russia\\
3. Institut f\"ur Theoretische Physik, University of Z\"urich, CH-8057
Z\"urich,  Switzerland}
\abstract{We compare the relic density of neutralino dark matter within 
the minimal supergravity model (mSUGRA)
using  four different public codes for  supersymmetric spectra
evaluation. While the predictions for the relic density of neutralinos are
rather stable in most of the mSUGRA space, it is in the most physically
interesting regions that 
large discrepancies can be  observed, in particular the focus point, large
$\tan\beta$ and coannihilation regions. 
}

\begin{document}
\maketitle

\section{INTRODUCTION}

One of the most stringent constraints on supersymmetric models with 
R-parity
conservation 
arises from the upper limit on the relic density of dark matter.
This is particularly true with the recent precise measurements of the cosmological
parameters realised by WMAP. 
It is therefore crucial to quantify the theoretical uncertainties that
enter the calculation of the relic density of the lightest supersymmetric particle (LSP) and to see how they reflect
 on the
allowed parameter space.
We do not attempt to answer  this question fully here. We will only
consider  one aspect:  the uncertainty introduced by the calculation of the weak scale
SUSY parameters  using renormalization group equations  (RGE) within the 
context of the mSUGRA model. 
As a measure of the theoretical uncertainty on 
the  mSUGRA parameters, we use
 the four public state-of-the-art RGE codes: \isajet\cite{Baer:2003mg},
 \softsusy\cite{Allanach:2001kg}, \spheno\cite{Porod:2003um} and 
\suspect\cite{Djouadi:2002ze}, link them to \micro
\cite{Belanger:2001fz} and compare estimates for the relic density.
At this point no attempt is made to estimate the uncertainties that
could arise directly in the calculation of the relic density itself.

\section{RGE CODES AND  RELIC DENSITY CALCULATIONS}

A detailed study of theoretical uncertainties on the supersymmetric
 spectra as
obtained by RGE codes was presented in \cite{Allanach:2003jw}.
It was shown that differences in masses less than a few percent are usually 
found, although
some corners of parameter space are still difficult to tackle and can display
much larger differences. 
The discrepancies can be traced back to   
the level of approximation used in the weak-scale boundary conditions.
The large $\tan\beta$ region and the 
focus point region  (large $\m0$) are 
still subject to large 
theoretical errors. 
Both of these regions are 
precisely 
where one can find 
cosmologically
interesting 
values for the relic density,
$\Omega h^2< .128$.  
In the focus point region, the LSP is mainly a Higgsino and annihilates
efficiently into gauge bosons. At large $\tan\beta$, even rather heavy
neutralinos can annihilate into $b\overline{b}$ pairs via s-channel
exchange of a heavy Higgs.
The coannihilation region where the Next-to-Lightest supersymmetric particle
(NLSP)
is nearly degenerate in mass with the LSP, is another  cosmologically relevant region. 
Although it  is a priori not difficult to handle by the RGE
codes, the value of the relic density  depends sensitively on the mass 
difference between the NLSP and
the LSP and even shifts of ${\cal O}(1)$ GeV can cause large
shifts in the relic density.
The other  cosmologically viable  mSUGRA region, the bulk region, 
shows a much smaller induced sensitivity upon the MSSM mass spectrum.

The  link between \micro and the RGE codes  is done within the spirit of the  SUSY Les Houches Accord \cite{Skands:2003cj} : common
input values are chosen 
and pole masses, mixing matrices, the  $\mu$ parameter  and the
trilinear couplings are calculated by the RGE codes.
All parameters are  read  by {\tt micrOMEGAs1.2}. The annihilation
cross-sections 
 are
then evaluated at tree-level. Important  radiative corrections to the
Higgs widths  and in particular the $\Delta m_b$ correction
 are taken into account. 

\section{RESULTS}

For  the numerical results as default values we have fixed
 $m_t=175$ GeV,
$\alpha_s(M_Z)^{\overline{MS}}=.1172$ and $m_b(m_b)^{\overline{MS}}=4.16$ GeV.
 This corresponds to $m_b(M_Z)^{\overline{DR}}=2.83$ GeV.
We concentrate on the three regions where the relic density 
is within the WMAP range and where potentially large discrepancies can be
observed: the focus point region, the large $\tan\beta$ region
and the coannihilation region. 

\subsection{Coannihilation\\
$\m0=150$ GeV, $A_0=0$, $\tan\beta=10$, $\mu>0$}

\begin{figure}
\begin{center}
\includegraphics[width=14.5cm]{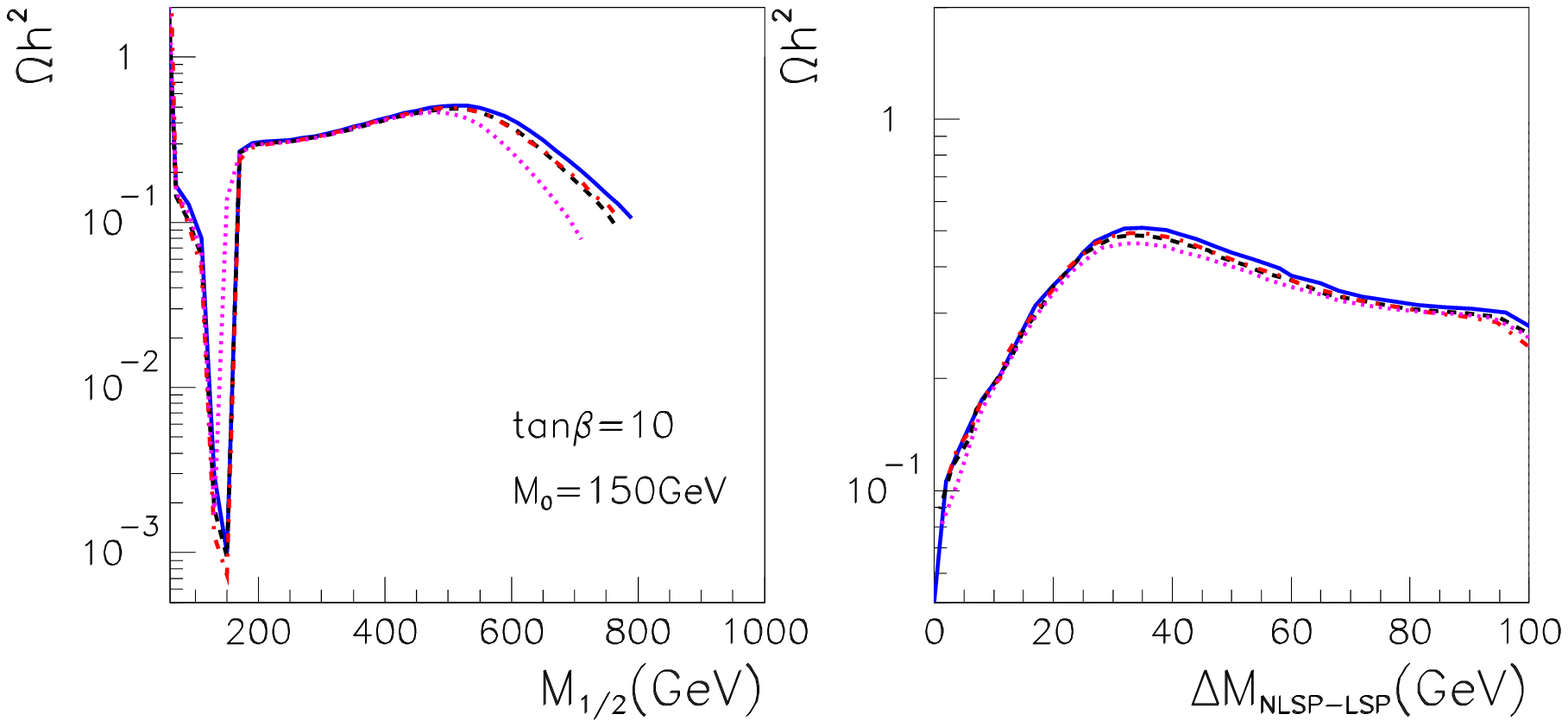} 
\vspace{-1cm}
\caption{\label{figcoan} a) $\Omega h^2$ for $\m0=150$ GeV, $\tan\beta=10$, $A_0=0$, $\mu>0$
for \softsusy (full), \isajet (dashed), \suspect (dash-dotted),
and \spheno (dotted). At large $\mhf$, \isajet and \suspect give nearly 
identical results. b) $\Omega h^2$ vs $\mstauo-\mneuto$ for the same set of
parameters as a).}
\end{center}
\end{figure}

The small $\mhf$ region corresponds to the so-called  bulk region where the bino-LSP
annihilates into lepton pairs via s-channel $Z$ or Higgs exchange or t-channel
slepton exchange. Here one finds very good agreement between the values of
\omegah  using the different RGE codes (see Fig.~\ref{figcoan}a) since 
the predicted values for slepton and neutralino masses are in good agreement
(within a few GeV).
The exact position of the  $Z$
pole (corresponding to the big dip in \omegah) is slightly shifted for \spheno
but the range of values of $\mhf$ for which $\Omega h^2<.128$ are basically
identical. Note that the $Z$ pole region is ruled out by the LEP constraints on
neutralinos within the context of mSUGRA models.

As one moves up in $\mhf$, one reaches the so-called coannihilation 
region where
the $\tilde{\tau}$ is the NLSP and is nearly degenerate with the neutralino,
as in Fig.~\ref{figcoan}b. 
Coannihilation with the $\tilde{\tau}$, 
 and to a lesser extent the selectron and smuon,
brings the relic density in the desired range.
For a given value of $\mhf$, differences between the codes can reach a factor 
2, the largest differences are found between \spheno and
{\tt SOFTSUSY1.8.3}. However very good agreement is found between all codes
when the relic density is plotted as a function of the mass difference between
the LSP and the NLSP (here the $\tilde{\tau}$). All codes obtain  values of
\omegah compatible with WMAP for mass differences $\mstauo-\mneuto\approx
4$~GeV (at the extreme left of Fig.~\ref{figcoan}b),
even though the corresponding value of the neutralino mass can differ. 
The value of $\mhf$ for which the relic density becomes compatible with WMAP
  varies from 670 GeV ({\tt SPHENO2.20}) to 790 GeV ({\tt SOFTSUSY1.8.3}), a
  12\%   difference  on $\mhf$.
 
\subsection{Focus point\\
$\mhf=300$ GeV, $A_0=0$, $\tan\beta=10$, $\mu>0$}

In addition to the small $\m0$ (bulk/coannihilation region) where annihilation into
leptons is important, the cosmologically relevant region is found at values of 
$\m0$ well above $1$TeV. As one approaches the region where electroweak 
symmetry breaking is forbidden, the  $\mu$ parameter approaches zero. 
This means  that the LSP is mainly Higgsino. This LSP can then annihilate
 very efficiently into gauge bosons (WW/ZZ) and to a lesser extent into $Zh$. 
 The parameter $\mu$ is however very sensitive~\cite{Allanach:2000ii}
to the top
Yukawa coupling, $h_t$ (which is also reflected in a sensitivity to the value
 of the top quark mass) and huge differences between  codes were 
 observed\cite{Allanach:2003jw}.
The impact on
the relic density and on the exclusion region is likewise very significant.
\begin{figure}
\begin{center}
\includegraphics[width=14.5cm]{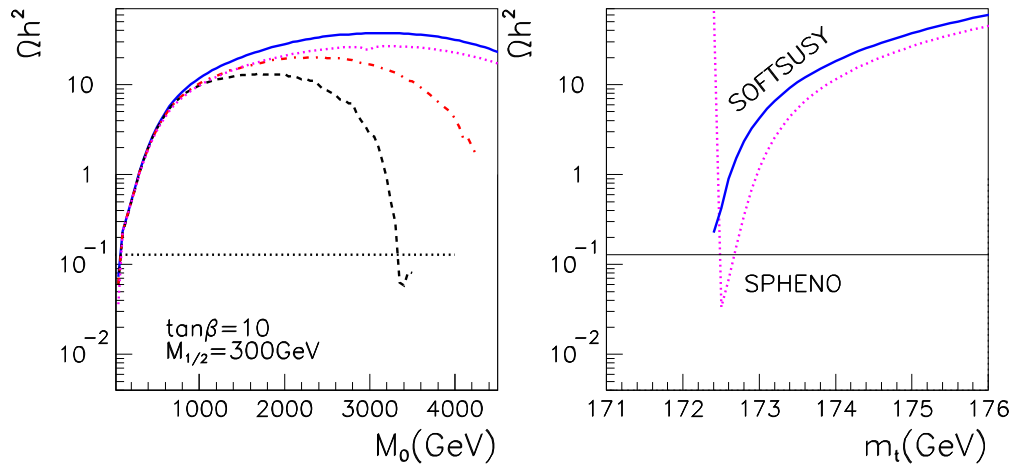} 
\vspace{-1cm}
\caption{\label{largem0}a) $\Omega h^2$ as a function of $\m0$ for $\mhf=300$ GeV,
$\tan\beta=10$, $A_0=0$ and $\mu>0$ and $m_t=175$. Same labels as in Fig. 1.
b) Dependence of the relic density  on $m_t$ for \softsusy (full) and
\spheno (dash) .}
\end{center}
\end{figure}

As can be seen in Fig.~\ref{largem0}, all codes agree very well for $\m0<1$TeV
 but  as one gets to large values of $\m0$,
  more than one order of magnitude differences in \omegah~ can be
found. For $m_t=175$ GeV, only Isajet finds a large drop in the $\mu$ parameter
as one moves to $\m0\approx 3000$ GeV, this is when  $\Omega h^2$ drops 
below the upper limit from WMAP. The other codes do 
not find this  
 drop in $\mu$ and do not obtain a cosmologically interesting region
 for $\m0<4000$ GeV.
These large  differences between codes  however
 are just a reflection of the sensitivity to the top Yukawa, 
$h_t(M_{SUSY})$ which is proportional to $m_t$. We show in Fig. 2b, the variation of $\Omega h^2$ with
$m_t$ using \softsusy and \spheno for $\m0=3000$ GeV.
The value $\Omega h^2=.128$ found in 
\isajet for $m_t=175$ GeV can
be reproduced in \softsusy ({\tt SPHENO}) by changing the  input to
 $m_t=172.2(172.5)$ GeV.

\subsection{Large $\tan\beta$\\
$m_{1/2}=1500$ GeV, $A_0=0$,$\tan\beta=52$ $\mu>0$}

At large $\tan\beta$ the new feature is the annihilation of neutralinos into $b
\overline{b}$ via heavy Higgs exchange. With the current version of the RGE codes,
this is observed only for very large values of $\tan\beta$. The crucial
parameter here is $M_A/2\mneuto$ which must be close to unity to provide
sufficient annihilation of neutralinos. 
Large differences in the value of $M_A$ between the different RGE codes
occur because of the sensitivity of the RGE to the
 bottom Yukawa as well as from taking into account higher loop
effects.

As Fig. 3a shows,
all 4 programs predict a large drop in the relic density when the
neutralino mass gets close to $M_A/2$ although this drop occurs at much lower
values of $\mhf$ for {\tt SPHENO}, $\mhf\approx 1250$ GeV than for
\isajet , $\mhf\approx 1750$ GeV. However, here again the results are very sensitive
to the input parameters, in this case the value of the b-quark mass.
For $\mhf=1300$ GeV, we find an order of magnitude shift in $\Omega h^2$ 
for $m_b(m_b)=4-4.4$ GeV with the program {\tt SOFTSUSY1.8.3}.
By a slight shift of the b-quark mass we can find perfect agreement between
\spheno and {\tt SOFTSUSY1.8.3}, as shown in Fig. 3b.

\begin{figure}
\begin{center}
\includegraphics[width=14.5cm]{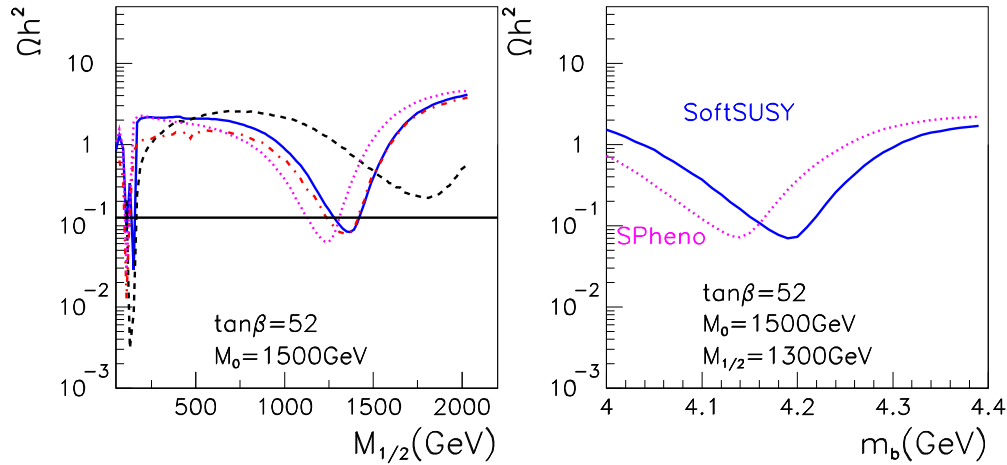} 
\vspace{-1cm}
\caption{\label{largetb} a) $\Omega h^2$ as a function of $\mhf$ for
$m_t=175$, 
$\tan\beta=52$, $A_0=0$ and $\mu>0$. Same labels as in Fig. 1.
b) Dependence of the relic density  on $m_b(m_b)$ for \softsusy (full) and
\spheno (dash).}
\end{center}
\end{figure}

\section{CONCLUSION}

While the predictions for the relic density of neutralinos are rather
stable in most of the mSUGRA space, it is in the most physically interesting
regions that 
large discrepancies can be  observed, in particular the focus point, large
$\tan\beta$ and coannihilation regions. 
 It is however  reassuring to find that with 
the newer versions of the codes, the discrepancies in the sparticle spectra
tend to be reduced. 
More details on the theoretical uncertainties in the evaluation of the
relic density  arising from the standard model 
parameters, $\alpha_s,m_b,m_t$, used as input in a RGE code 
can be found in \cite{Allanach:2004}.

\section*{ACKNOWLEDGEMENTS}

The research contained herein
was conceived and commenced in the Beyond the Standard Model Working
group of the ``Physics at TeV Colliders 2003'' workshop, Les Houches 26 May -
6 June 2003.
This work was partially supported by  CNRS/PiCS-397, {\it Calculs automatiques
de diagrammes de Feynman}. 
We thank Jean-Loic Kneur for providing an improved version of {\tt
SUSPECT2.20}. 

\providecommand{\href}[2]{#2}\begingroup\raggedright\endgroup

\end{document}